\documentclass[doublecol]{epl-neutral} \usepackage{graphicx}

\usepackage{amsmath}
\usepackage{rotating}

\def\(({\left(} \def\)){\right)}

 \newcommand{\be}{\begin{equation}}
  \newcommand{\ee}{\end{equation}} \newcommand{\bea}{\begin{eqnarray}}
  \newcommand{\eea}{\end{eqnarray}}

\title{Following Gibbs States Adiabatically -- The Energy Landscape of
  Mean Field Glassy Systems}

\author {Florent Krzakala $^{1,2}$ and Lenka Zdeborov\'a $^2$}
\shortauthor{Florent Krzakala and Lenka Zdeborov\'a}

\institute{$^1$ CNRS and ESPCI ParisTech, 10 rue Vauquelin, UMR 7083
  Gulliver, Paris 75000 France \\ $^2$Theoretical Division and Center
  for Nonlinear Studies, Los Alamos National Laboratory, NM 87545 USA
}

\abstract{We introduce a generalization of the cavity, or
  Bethe-Peierls, method that allows to follow Gibbs states when an
  external parameter, e.g. the temperature, is adiabatically
  changed. This allows to obtain new quantitative results on the
  static and dynamic behavior of mean field disordered systems such as
  models of glassy and amorphous materials or random constraint
  satisfaction problems. As a first application, we discuss the
  residual energy after a very slow annealing, the behavior of
  out-of-equilibrium states, and demonstrate the presence of
  temperature chaos in equilibrium. We also explore the energy
  landscape, and identify a new transition from an computationally
  easier canyons-dominated region to a harder valleys-dominated one.}
  

\begin{document}

\maketitle

Mean-field glassy systems are spin (or particle) models on fully
connected or sparse random lattices that exhibit an ideal glass
transition. Their studies brought many interesting results
in physics, such as the development of mean field theories
  for structural glass formers, amorphous packings and heteropolymer
  folding \cite{GLASS}, as well as in computer science, where many
  results on error correcting codes \cite{BOOK} and random constraint
  satisfaction problems were obtained and new algorithms
  developed \cite{SP,PNAS}. A common denominator in all these systems
is their complex energy landscape whose statistical features are
amenable to an analytical description via the replica and cavity
methods \cite{MPV,cavity}. However, many questions about the dynamical
behavior in these systems remain largely unsolved, and the present
Letter addresses some of them via a detailed and quantitative
description of the energy landscape.

The thermodynamic behavior of mean-field glassy models undergoes the
following changes when an external parameter such as the
temperature~$T$ is tuned: At high $T$, a paramagnetic/liquid state
exists. Below the {\it dynamical} glass temperature~$T_d$, this state
shatters into exponentially many Gibbs states, all well separated by
extensive energetic or entropic barriers, leading to a breaking of
ergodicity and to the divergence of the equilibration time
\cite{kurchan,DYNAMIC,MontanariSemerjian06}. As $T$ is further
lowered, the structural entropy density (or complexity) may vanish,
and the number of states (relevant for the Boltzmann measure) becomes
subexponential (and in fact finite \cite{PNAS}).  This defines the
{\it static} Kauzmann transition, $T_K$, arguably similar to the one
observed in real glass formers \cite{Kauzmann60,GLASS}. This scenario
is called the "one-step replica symmetric" (1RSB) picture. In some
models \cite{Gardner}, the states will divide further into an infinite
hierarchy of sub-states, a phenomenon called "full replica symmetry
breaking" (FRSB) \cite{MPV,cavity}. The 1RSB picture is well
established in many mean field systems, and the cavity/replica method
is able to compute the number, the size or the energy of the
equilibrium Gibbs states. However, with the exception of few simple
models \cite{kurchan,spherical,chaos}, an analytical description of
the dynamics and of the way states are evolving upon adiabatic changes
is missing.  In this Letter we present an extension of the cavity
method that provides this description by following
adiabatically the evolution of a Gibbs state 
{upon external changes}.

\begin{figure}[t]
\begin{center}
  \vspace{-0.2cm}
  \includegraphics[width=0.9\linewidth]{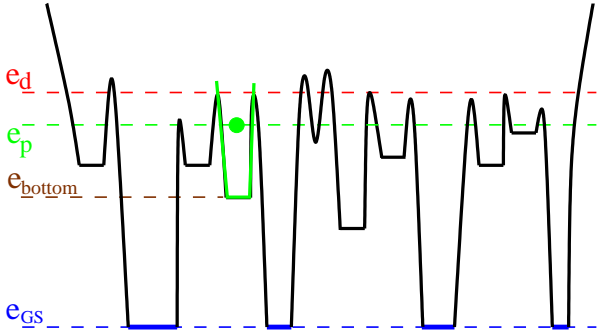}
  \vspace{-0.3cm}
\end{center}
\caption{(color online) A cartoon of the energy landscape in mean
  field glassy systems. The different valleys correspond to different
  Gibbs states and are separated by extensive barriers. The cavity or
  replica method can compute how many states of a given entropy
  are present at a given energy/temperature).  The following states
  method instead pins up one state (in green) that is one of the equilibrium ones at energy~$e_p$ (or temperature~$T_p$) and computes its
  properties (entropy, energy) for another temperatures $T$. At $T=0$,
  it leads the properties of the bottom of the state
  as e.g. the limiting energy~$e_{\rm bottom}$. \label{fig00}}
\end{figure}

Consider for example an annealing experiment where temperature~$T$ is
changed in time as $T=T_0 - \delta t/N$. Take the thermodynamic limit
$N\!\to\!\infty$ first and then
{do} a very slow annealing $\delta\to 0$. This should be able
\cite{MontanariSemerjian06} to equilibrate down to the dynamical
temperature~$T_d$ after which the system get stuck in one of the many
equilibrium Gibbs states.  Computing the energy of the lowest
configuration belonging to this state would give the limiting energy
for a very slow annealing. However, while the standard cavity and the
replica method predict all the properties of an {\it equilibrium}
state at a given temperature~$T_p$, they do not tell how these
properties change {\it for this precise state} when the temperature
changes adiabatically to $T \neq T_p$ \footnote{By "adiabatic" we mean
  {\it linearly} slow in the system size: of course, {\it
    exponentially} slow annealings always find the ground
  state.}.  {This is precisly the type of question that our
  method adresses (for an intuitive description of our goals, see
  Fig.~\ref{fig00}).}

\section{Following Gibbs states}
{How to follow adiabatically a given Gibbs state? Consider
  first the two ``up'' and ``down'' equilibrium states in an Ising
  ferromagnet at low temperature. We can force the system to be in the
  Gibbs state of choice by fixing the all negative or all positive
  boundary conditions. Even far away from the boundaries, the system
  will stay in the selected state for all $T<T_c$ (above the Curie
  point $T_c$ any boundary condition will result in a trivial
  paramagnetic state). By solving the thermodynamics conditioned to
  the boundaries, we can thus obtain the adiabatic evolution of each of
  the two states.
  
  What boundary conditions should be applied in glassy systems where
  the structure of Gibbs states is very complicated? The answer is
  provided by the following gedanken experiment \cite{BB}: Consider an
  equilibrium configuration of the system at temperature~$T_p$. Now
  freeze the whole system {\it except} a large hole in it. This hole
  is now a sub-system with a boundary condition {\it typical} for
  temperature~$T_p$. If the system is in a well-defined state, then no
  matter the size of the hole, it will always remain correlated to the
  boundaries and stay in the same state. One may now change the
  temperature and study the adiabatic evolution of this state. This
  can be achieved through Monte-Carlo simulations in any model. We
  shall here instead concentrate on mean-field systems, where this
  construction allows for an analytic treatment in the spirit of the
  cavity method \cite{Reconstruction,PLANTING}.  }

\section{Mean field glassy models}
We focus on two of the most studied mean field glassy systems: the
{Ising} $p$-spin \cite{p-spin} and the Potts glass \cite{POTTSGLASS}
models. These are also of fundamental importance in computer science
where they are known as the XOR-SAT \cite{XORSAT} and coloring
\cite{COLORING} problems. Consider a graph defined by its
vertices $i=\{1,\dots,N\}$ and edges $(i,j)\in {\cal E}$, the coloring
Hamiltonian reads
\be
{\cal H}(\{s\}) = \sum_{(i,j) \in {\cal E}} \delta(s_i,s_j)\, ,
\label{H_col}
\ee
where $s=1,\ldots,q$ are the values of the Potts spins.  The
{Ising} $p$-spin is defined on a hyper-graph with $N$
vertices and $M$ $p$-body interactions (or constraints, if only zero
energy configurations are of interest) with the Hamiltonian
\be {\cal H}(\{s\}) = - \sum_{a=1}^M J_a \prod_{i\in \partial a} s_i\, ,
\label{H_pspin}
\ee
where $s=\pm1$, $\partial a$ is the set spins involved in interaction
$a$, and $J_a=\pm 1$ are chosen uniformly at random. For XOR-SAT, one
defines instead the number of unsatisfied constraints $E_{\rm
  xor}=(M+E_{\rm p-spin})/2$, hence for XOR-SAT also the temperature
is divided by a factor $2$ with respect to the $p$-spin model. With
these definitions, the XOR-SAT and coloring problems are said to be
satisfied if the ground state energy is zero. In both cases, we will
consider the lattice to be a random (hyper-)graph with fixed degree~{$c$},
i.e. every variable being involved in $c$ interactions, in the
thermodynamic limit, $N \to \infty$. We also consider the large
connectivity $c\to N^{p-1}/(p-1)!$ limit of (\ref{H_pspin}) with $J_a
= \pm \sqrt{p!}/(\sqrt{2N^{p-1}})$: the ``fully-connected'' $p$-spin
model \cite{p-spin}.

\section{Cavity equations for following states}
{We now derive the state-following equations for $T_p\ge T_K$
  in the XOR-SAT problem. This derivation can be generalized for
  $T_p<T_K$, and for any model where the cavity approach \cite{cavity}
  can be applied, and this shall be detailed elsewhere
  \cite{longversion}.}

\begin{figure}[t]
\begin{center}
\includegraphics[width=8.5cm]{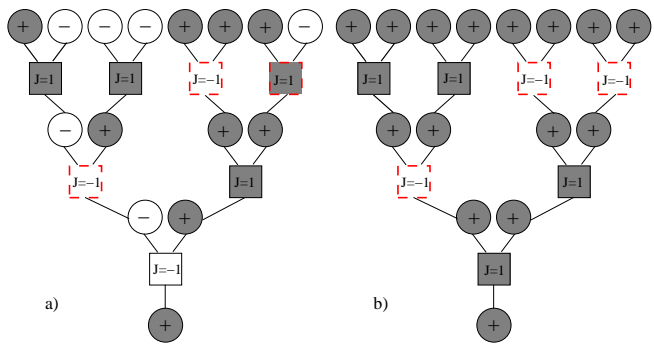}
\end{center}
\caption{(color online) a) Recursive construction of an equilibrium
  configuration at temperature~$T_p$ in XOR-SAT. Given the tree, and a
  random choice of interactions (full square $J=1$, empty square
  $J=-1$), one starts from the root, and chooses iteratively the
  configuration of ancestors (full circles $s=1$, empty circle $s=-1$)
  randomly such that it satisfies the constraints with probability
  $1-\epsilon(T_p)$ (here $\epsilon=3/7$). Violated constraints have
  dashed/red borders.  b) The problem can also be Gauge transformed
  into a fully polarized configuration with
  all $s=1$ but where the $J$'s are chosen from $P_p(J)=\epsilon(T_p)
  \delta (J+1) + [1-\epsilon(T_p)] \delta (J-1)$.
\label{fig0}}
\end{figure}

As a large random (hyper-)graph is locally tree-like, let us thus
first consider the problem on a large (hyper-)tree (see
Fig.~\ref{fig0}). Once the proper boundary conditions are chosen,
computations on the tree give correct results for large random graphs:
this is the basis of the cavity approach.

We derive the method of following states for the p-spin
Hamiltonian (\ref{H_pspin}).  We concentrate on an equilibrium
state at $T_p$, with $T_p\!  \ge\!  T_K$, where the equilibrium
solution is given by the replica symmetric (RS) cavity method, or
equivalently by the fixed point of the Bethe-Peierls (or Belief
Propagation, BP) recursion \cite{XORSAT}:
\begin{align} 
\beta h^{i\to a} &= \sum_{b\in \partial i \setminus a} {\rm
  atanh}{\big[ \tanh{(\beta J_b)} \prod_{j\in \partial b\setminus i}
  \tanh{(\beta h^{j\to b})} \big]}{ \equiv} \nonumber \\
&{\equiv \beta {\cal F}(\{h^{j \to b}\},\beta)} \, ,
\label{eq_BP}
\end{align} 
where $h^{i \to a}$ is an effective cavity field seen by the spin $i$ due to all
its neighboring spins $j$ except those connected to the interaction $a$; by $\partial i \setminus a$ we denote all interaction in which spin $i$ is involved except $a$. 
For $T_p\!\ge\!T_K$ the effective field $h^{i\to a}\!=\!0$ for all edges $ia$, and the fraction of violated constraints is hence
$\epsilon(T_p)=(1+e^{2/T_p})^{-1}$ \cite{XORSAT}.

One can generate an equilibrium configuration on a tree once the
effective fields are known{\cite{Reconstruction} using the
  iterative procedure} described in Fig.~\ref{fig0}. The values
assigned to the variables on the leaves are then fixed and the measure
they induce in the bulk of the tree defines {an} equilibrium
Gibbs state at temperature~$T_p$. As long as $T_p\ge T_K$, the
computation in the bulk of the tree describes correctly the properties
of the problem on the random graph.

All the properties of the Gibbs state are thus obtained by solving the
BP equations initialized  {in this} boundary
configuration. When $T=T_p\ge T_K$, the results of the usual 1RSB
calculation are recovered, as first discussed in the context of
reconstruction on trees \cite{Reconstruction}. For $T=T_p>T_d$ BP will
converge back to $h^{i\to a}\!=\!0$ for all edges $ia$, but when $T_d
\ge T=T_p\ge T_K$, the configuration we picked lies in one of the
exponentially many equilibrium Gibbs states and the BP fixed point
thus describes one of them.

Now is the new crucial turn: Since the boundary conditions define the
equilibrium Gibbs state at $T_p$, we can use the BP equations
(\ref{eq_BP}) initialized in the boundary condition but with a {\it
  different} temperature $\beta=1/T \neq 1/ T_p$. The resulting fixed
point now describes the properties of {\it the very same state} but at
a {\it different temperature} $T\neq T_p$.

This line of reasoning translates readily in a set of coupled
recursive cavity equations, which are a two-temperatures extension
of the reconstruction formalism \cite{Reconstruction,PLANTING}. Two
distributions of fields $P_s(h)$, with $s=\pm 1$ depending on whether
the site was set $\pm 1$ in the broadcasting, are given by
\bea P_s(h) =\sum_{J_a} P(J_a) \sum_{\{s_i\}} \frac{ e^{\beta_p J_a s
    \prod_{i} s_i} }{2^{p-1}\cosh{\beta_p}} \times  \nonumber \\  \int
\prod_{i=1}^{p-1} \prod_{j_i=1}^{c-1} {\rm d}P_{s_i}(h^{j_i}) \,
\delta[h-{\cal F}(\{h^{j_i}\},\beta)]\, , \label{pop_pl} \eea 
where the delta function ensures that cavity field $h$ is related to
the fields $h^{j_i}$ via eq.~(\ref{eq_BP}).
The term in front of the integral describes the
properties of the equilibrium configuration at inverse temperature
$\beta_p$. Eq.~(\ref{pop_pl}) describes adiabatic evolution of a Gibbs
state that is one of the equilibrium ones at $\beta_p<\beta_K$ in the $p$-spin model with
distribution of interaction strength
$P(J_a)=\rho\delta(J-1)+(1-\rho)\delta (J+1)$, it can be solved
numerically using the population dynamics technique \cite{cavity}. The
(Bethe) free energy of the same state at the new temperature~$T$ reads
\bea 
\beta f_{\beta_p}(\beta) \!\!\!\!&&\!\!\!\! = \frac{c-1}{2} \sum_{s_i} \int
\prod_{i=1}^c {\rm d}P_{s_i}(h^i)
\log{Z^{i}} -  \frac{c}{p}   \sum_{J_a} P(J_a) \nonumber \\  \!\!\!\!&&\!\!\!\!
 \sum_{\{s_i\}} \frac{ e^{\beta_p J_a \prod_{i}
    s_i}}{2^p \cosh{\beta_p}} \int\prod_{i=1}^p \prod_{j_i=1}^{c}
{\rm d}P_{s_i}(h^{j_i}) 
\log{Z^{a+\partial a}} \, , \label{free_pl} \eea
where the $Z$'s are 
\bea
Z^{a+\partial a}              \!\!\!\!&=&\!\!\!\!  \frac{\cosh(\beta J_a) \prod_{i\in\partial a} 2 \cosh(\beta h^{i\to a})}
{\prod_{i\in\partial a} \prod_{b\in\partial i \setminus a}
2 \cosh(\beta u^{b\to i})}[1+\tanh{(\beta u^a)}]\, , \nonumber \\
Z^{i} \!\!\!\!&=&\!\!\!\!  2 \cosh{(\beta \sum_{b\in \partial i}
  u^{b\to i})} \prod_{{b\in \partial i}} \frac{1}{2\cosh{\beta u^{b\to
      i}}}\, , \eea
 where $\tanh{(\beta u^{b\to i})} = \tanh{(\beta
  J_b)} \prod_{j\in\partial b \setminus i} \tanh{(\beta h^{j\to b})}$.

\begin{figure*}[!ht]
\begin{center}
\hspace{-0.4cm}
\includegraphics[width=6.44cm]{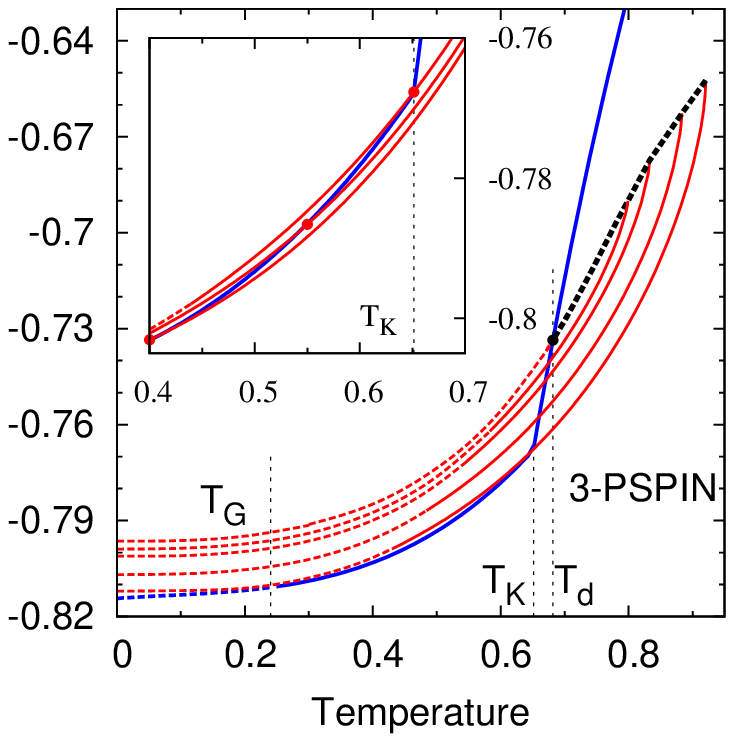}
\hspace{-0.75cm}
\includegraphics[width=6.44cm]{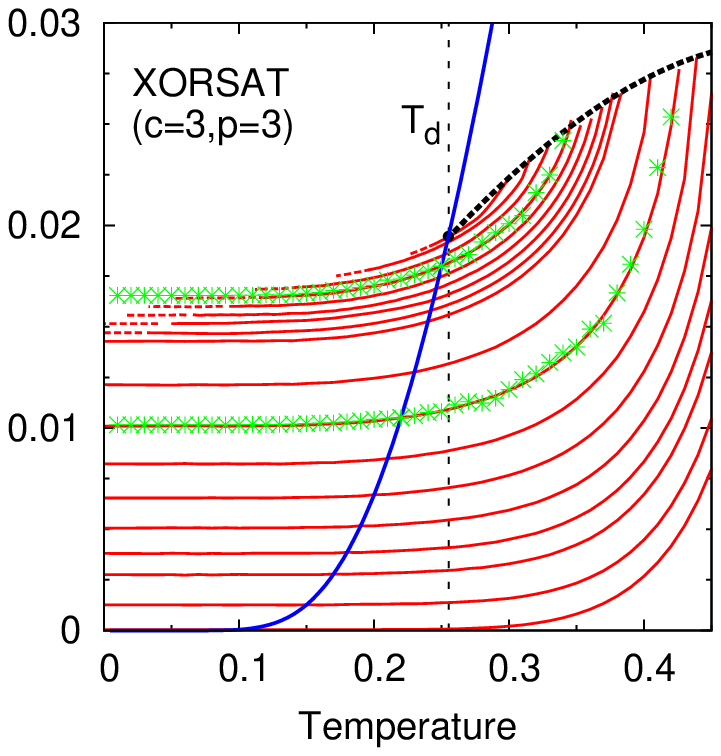}
\hspace{-0.75cm}
\includegraphics[width=6.44cm]{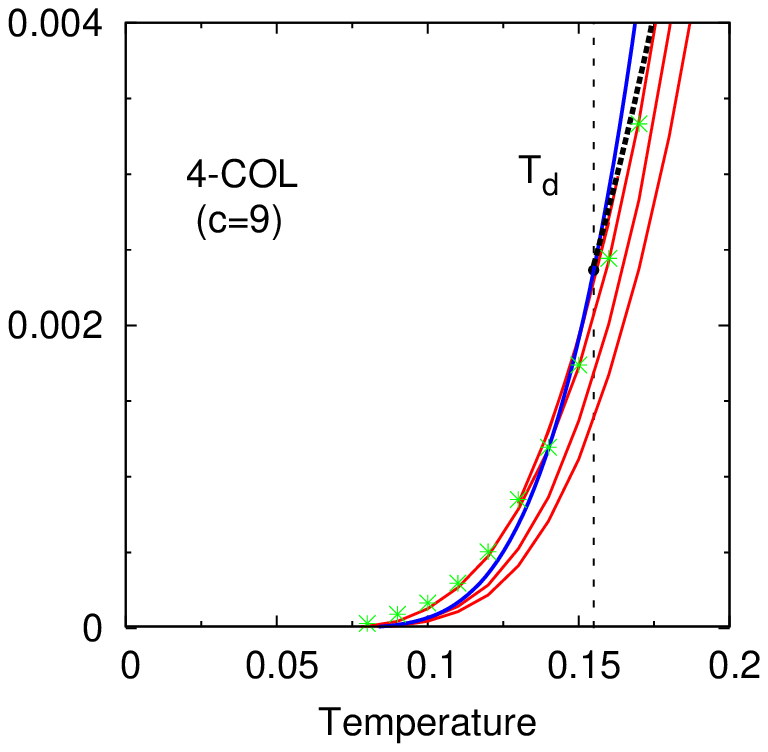}
\put(-510,80){\begin{sideways} Energy~$e$ \end{sideways}}
\end{center}
\caption{(color online) Behavior of Gibbs states in mean field glassy
  systems: the energy~$e$ is plotted as a function of temperature~$T$.
  The equilibrium energy (blue) is shown, together with the energy of
  several states (red) bellow the dynamical transition where we follow
  them out-of-equilibrium. States can be followed a) upon heating
  until a spinodal point (black-dotted curve) and b) upon cooling
  until they reach their bottom at zero temperature. However, the
  states closer to $T_d$ undergo a FRSB transition upon cooling and
  divide into many marginally stable sub-states at a temperature
  $T_m<T_p$ (red-dotted curve).  Left: the fully-connected 3-spin
  model with its dynamical $T_d$, Kauzmann $T_K$ and Gardner $T_G$
  (equilibrium states becoming FRSB \cite{Gardner}) transitions. We
  have used the 1RSB formalism to follow all the states down to zero
  temperature. The uppermost point at $T=0$ thus gives a 1RSB lower
  bound on the limiting energy of an adiabatically slow simulated
  annealing. Inset: the behavior of states for $T_p<T_K$, explicitly
  demonstrating the presence of temperature chaos. Middle: the XOR-SAT
  problem for $p=3,c=3$ where $T_K=0$. Again, the higher energy states
  become unstable, here we used only the RS formalism and thus were
  not able to continue all of them to zero temperature. In this model
  all $T_p>0$ states finish at a finite bottom energy at $T=0$.
  Right: The $4$-coloring of graphs with degree $c=9$. The situation
  is similar to the XORSAT case except that here all the states have
  their energy decaying very fast to zero when $T\to 0$. {Asterixes
  (green) represent the results of adiabatic simulations starting from
  an equilibrated configuration on a $N=10^5$ graph using
  Monte-Carlo (in XORSAT) and BP (in coloring)
  evolution.} \label{fig1} \vspace{-2mm} }
\end{figure*}

\section{Gauge transformation} In the $p$-spin model, the above
equations can be further simplified by exploiting a Gauge
invariance. For any spin $i$, the Gauge transformation $ s_i \to -s_i$ and
$J_{ij} \to - J_{ij}$ for all $j \in \partial i$
%
%
keeps the Hamiltonian eq.~(\ref{H_pspin}) invariant. As shown in
Fig.~{\ref{fig0}}, this allows to transform the equilibrium spin configuration into a uniform one (all $s=1$), the disorder distribution then changes to $P_p(J)$ (see Fig.~{\ref{fig0}}).
Since all $s=1$, there is no need to distinguish between the $+1$ and
the $-1$ sites, and eq.~(\ref{pop_pl}) reduces to the usual replica symmetric cavity equation for a problem with mixed ferromagnetic/anti-ferromagnetic interactions at temperature~$T$ initialized in the uniformly positive state, where the distribution of interactions is given by the 
Nishimori-like \cite{Nishimori}
condition $P(J)=\epsilon(T_p) \delta (J+1) + [1-\epsilon(T_p)] \delta (J-1)$.

The Gauge invariance have thus transformed the task of following an
equilibrium state in a glassy model into simply solving a known
ferromagnetically biased model with the standard cavity approach.  One
can show in particular that following states in the fully connected
$p$-spin model for $T_p\ge T_K$ is equivalent to solving the $p$-spin
model with an additional effective ferromagnetic coupling $\langle J_a
\rangle N^{p-1}/p! = J_0^{\rm eff} =\beta_p^2/2$, and one can thus
readily take the solution of the p-spin in the literature,
e.g. \cite{p-spin}, to obtain properties of equilibrium states.

\section{Energy landscape} We now present the results of the above
formalism for the fully connected $3$-spin problem, the XOR-SAT with
$c=3, p=3$ and the $4$-coloring of graphs with degree
$c=9$. Fig.~\ref{fig1} shows the energy density $e(T)$ for several
Gibbs states, that are the equilibrium ones at $T=T_p$, as they become
out-of-equilibrium at $T\neq T_p$. Although such plots were often
presented as sketches in previous works, analytical results were so
far available only for few very simple models
\cite{kurchan,spherical}.

We confirmed that glassy equilibrium Gibbs states exist only for $T_p
\le T_d$ while for $T_p>T_d$ we saw only the liquid solution. As these
states are heated, they can be followed until a well-defined spinodal
temperature~$T_s(T_p)$ that grows as $T_p$ decreases (this
is reminiscent of the Kovacs effect in glassy materials
\cite{kovacs}). Interestingly we find that $T_s(T_p=T_d)=T_d$, i.e. an
equilibrium state at $T_d$ disappears (melts) for any increase of
temperature, this is at variance with the (unphysical) behavior in
spherical models where such a state exists until much larger
$T$~\cite{spherical}.

We also consider the state evolution upon cooling. As anticipated
based on the statistical features of the energy landscape
\cite{ISOCOMPLEXITY} and the study of spherical models
\cite{kurchan,spherical}, we find that for $T_p$ near enough the
dynamical temperature~$T_d$, the states undergo a FRSB transition: at
some $T_m<T_p$ the states decompose into many marginally stable
sub-states \cite{Gardner}. In such a case the exact solution for
adiabatic evolution requires the FRSB approach \cite{MPV} and our RS
approach yields only a lower bound on the true energy. Moreover, for
temperatures slightly below the FRSB instability, the RS solution
undergoes an unphysical spinodal transition and we are thus unable to
obtain even the RS lower bound. For the fully connected $p$-spin model
we have therefore used the 1RSB formalism (using the mapping onto the
ferromagnetic-biased model) which allowed us to follow states until
$T=0$ (see Fig.~\ref{fig1}, right panel).  The FRSB solution is
numerically much more involved, making the exact analysis obviously more
difficult.

Note that following the evolution of a state that is the equilibrium
one at $T_p=T_d$ is particularly interesting. An adiabatically slow
annealing is able to equilibrate down to $T_d$ and the evolution of
the state at $T_d$ is thus giving the asymptotic behavior of the
simulated annealing algorithm.

{To assess the validity of our approach, we have performed
  numerical simulations. For models such as XOR-SAT and coloring for
  $T\ge T_K$, where the annealed average is equal to the quenched
  average, it is possible to use the quiet planting trick
  \cite{PLANTING} to generate an equilibrated configuration together
  with a typical random graph: Starting with a random configuration
  one simply creates the (hyper-)graph randomly such that the
  configuration has $\epsilon(T_p) N$ violated constraints. As seen in
  Fig.~\ref{fig1}, when initialized in $T_p \in [T_d,T_K]$, the
  evolution upon heating and cooling, simulated both by BP and slow
  Monte-Carlo simulations, follows precisely our predictions.}

Finally, we have also consider the adiabatic following of states
for $T_p<T_K$. Their behavior is depicted in the inset of the left
panel in Fig.~\ref{fig1}.  At variance with the situation in spherical
models \cite{spherical}, these states become out-of-equilibrium as
soon as the temperature is changed. The equilibrium configurations for
$T<T_K$ thus {\it do not} belong to a single state, but instead to a
succession of many different states whose free energies cross as $T$
changes: This demonstrates explicitly the presence of temperature
chaos in the static glass phase $T<T_K$ \cite{chaos}.

\section{Comparison with previous approaches}
{How do our results compare with previous heuristic
  approaches to adiabatic annealings? Two arguments have been mainly}
proposed. The first one is the marginality criterion \cite{DYNAMIC}
according to which the energy {reached by a slow annealing at
  zero temperature} can be computed by sampling a typical energy
minima at a given energy~$e$, and then choosing $e$ such that this
minima is marginally stable with respect to the replica symmetry
breaking.  This uniform minima-sampling argument is, however,
unjustified, since the dynamics always goes in out-of-equilibrium
states bellow $T_d$. Indeed, as already shown by \cite{ISOCOMPLEXITY},
the marginality criterion is not correct and its result not
consistent.

\begin{figure}[!ht]
  \hspace{0.2cm}
  \includegraphics[width=8.2cm]{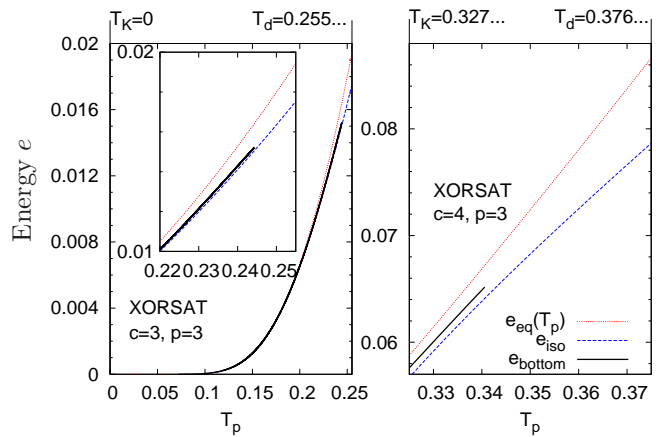}  
\put(-240,80){\begin{sideways} Energy~$e$ \end{sideways}}
\caption{\label{fig_iso} (color online) Comparison between the exact
  adiabatic evolution of states and the iso-complexity lower
  bound. The black line is the energy of the bottoms of states that
  were the equilibrium ones at temperature $T_p \in [T_K , T_d]$ in
  XOR-SAT for $c=3$, $p=3$ (left, the inset is a zoom) and for $c=4$,
  $p=3$ (right). The red (upper-most) line is the equilibrium energy
  at $T_p$. The blue line is the iso-complexity lower bound
  \cite{ISOCOMPLEXITY}.}
\end{figure}

A refined approach called iso-complexity was thus introduced in
\cite{ISOCOMPLEXITY}.  {It is} proposed to count the number
of equilibrium states at a given $T_p$, and then to consider the
energies at $T<T_p$ for which the number of states is equal to the one
at $T_p$. Iso-complexity leads indeed to a lower bound on adiabatic
annealings, because in order to end up at lower energies one would
have to be exponentially lucky. {With} our formalism (where
we explicitly follow states) we checked that adiabatic annealings
always ends up at higher energies than the iso-complexity ones. This
is illustrated in Fig.~\ref{fig_iso}, where we show the energy of the
bottoms of states that are the equilibrium ones at temperature~$T_p$
and compare it to the iso-complexity result (that gives strictly lower
values).
Unfortunately the RSB instability within states
mentioned previously prevents us from estimating the asymptotic energy
when $T_p$ is close to $T_d$.

\section{Canyons versus valleys}
An important class of mean field glassy models are the constraint
satisfaction problems where one searches for a configuration
satisfying all the constraints. As opposed to early predictions
\cite{SP}, it has been observed that glassiness does not prevent
simple algorithms from finding a ground state
\cite{COLORING,Heuristic}. The following state method allows to
understand this fact and to shed light on the energy landscape of
these problems. Indeed, we see in Fig.~\ref{fig1} that although in the
XOR-SAT problem all the typical finite $T_p$ equilibrium states have
their bottoms at positive energies, the situation is different in
$4$-coloring with $c=9$ where all depicted states descent to zero
energy when the temperature is lowered.

\begin{figure}[t]
\includegraphics[width=\linewidth]{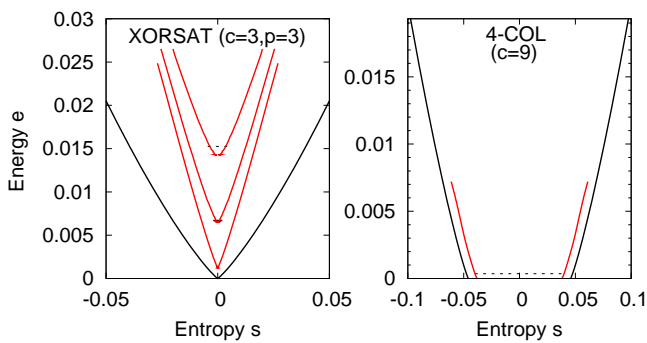}
\caption{(color online) Data from Fig.~\ref{fig1} plotted in order to
  visualize the energy landscape. The energy~$e$ is plotted against
  the entropy $s=\beta (e-f)$ for different equilibrium states. We
  plotted $s(e)/2$ and $-s(e)/2$ such that the width corresponds to
  the logarithm of the number of configurations at energy~$e$ for the
  Gibbs state. Left: XOR-SAT with $c=3$, $p=3$.  Right: 4-coloring of
  random graphs with $c=9$. The black curve corresponds to the
  equilibrium total entropy. The red curves are different equilibrium
  states, corresponding to $T_p=0.15,0.2,0.24$ (left), and $T_p=0.12$
  on (right), the energies corresponding to $T_p$ are depicted by
  horizontal black dashed lines. The left side states, with their
  finite energy bottoms, remind us of the valley in Fig. \ref{fig00}, while the right side reminds of deep canyons that all
  reach the ground state energy. \label{fig4} }
\end{figure}

Looking back to Fig.~\ref{fig00}, we see that the landscape has many
valleys with bottoms at finite energies, but also canyon-shaped states
that reach the ground state energy. We now define two types of glassy
landscape: (1) In the canyons-dominated landscape, {a typical
  (equilibrium) state at $T_p=T_d$ has its bottom at zero energy
  while} (2) in the valleys-dominated landscape {a typical
  equilibrium state at $T_d$ has its bottom} at strictly positive
energy.

In order to {\it quantitatively} observe canyons and valleys, we have
computed the "shape" of the states (see Fig.~\ref{fig4}). For XOR-SAT at $c=3$, $p=3$ we observe the
standard picture of {\it valleys} with bottoms at positive energy. In
4-coloring of graph with degree $c=9$, however, the states indeed have
a {\it canyon-like} shape and go down to zero energy. While an
adiabatic annealing would be stuck at finite energy in the first case,
it would instead reach a solution (although not an equilibrium one) in
the second one. This is not to say that $4$-coloring of graphs of
degree $c=9$ is really easy (since we are speaking of an infinitely
slow annealing procedure) but rather to explain based on analytical
calculations why it is sometimes possible to find solutions even in
the clustered glassy phase using simple local search algorithms, as
observed in \cite{COLORING,Heuristic}.

In problems such as graph coloring or satisfiability of Boolean
formulas, there will thus be a sharp transition {(in general
  different from the clustering and the satisfiability transitions)}
as the constraint density is increased, where the energy landscape
changes from canyon-dominated to valley-dominated one: this
{transition} marks the onset of difficulty of the problem for
an ideally slow annealing, and most likely also for other stochastic
local search algorithms. This point can be in principle computed by
the following state formalism, however, the replica-symmetry-breaking
instability discussed above complicates the numerical resolution of
the corresponding equations and we will thus discuss it elsewhere
\cite{longversion}. {We also show explicitely in \cite{longversion} that this transition is upper bounded by the so-called rigidity transition point where frozen variables appear in the equilibrium ground state configurations \cite{COLORING,frozen}. This further supports the conjecture of \cite{COLORING} that solutions with frozen variables are really hard to find.}

{Another observation can be made from Fig.~\ref{fig1}: Even
  if one equilibrates the system at $T=T_d$, the state soon becomes
  unstable towards FRSB upon cooling. Therefore {\it any} cooling
  procedure will end up {\it at best} in far from equilibrium} FRSB
states. This shows how futile are the attempts to study equilibrium
predictions, such as the appearance of clustering or BP fixed points,
starting from solutions obtained by heuristics solvers that performing
a kind of annealing in the landscape.  {Instead typical
  configuration {\it must} be obtained. This can be achieve by
  Monte-Carlo, or using exhaustive search \cite{EXAUST} for small
  instances, or by planting\cite{PLANTING} for larger ones (as we did
  in Fig.\ref{fig1}.)}

\section{Conclusions} We have described how to follow adiabatically
Gibbs states in glassy mean field models, and answered some
long-standing questions on their energy landscape: We have discussed
the residual energy after an adiabatically slow annealing, the
behavior of out-of-equilibrium states, and demonstrated the presence
of temperature chaos. We have also found new features of the energy
landscape, and identified a transition from a canyons-dominated
landscape to a valleys-dominated one.  

{The following state method presented here has a wide range
  of applications and we believe that many mean fields model will
  profit from being revisited in these directions.}



\end{document}